\documentclass[]{elsarticle}

\usepackage[utf8]{inputenc}
\usepackage{graphicx}
\usepackage{subfigure}

\usepackage{amssymb}
\usepackage{amsthm}
\usepackage{url}
\usepackage{amsmath}
\usepackage{subfigure}
\usepackage{textcomp}
\usepackage{xcolor}

\graphicspath{{figs/}}

\title{Fungal sensing skin}

\author[1]{Andrew Adamatzky}
\author[2]{Antoni Gandia}
\author[3,1]{Alessandro Chiolerio}

\address[1]{Unconventional Computing Laboratory, UWE, Bristol, UK}
\address[2]{Mogu S.r.l., Inarzo, Italy}
\address[3]{Center for Sustainable Future Technologies,
Istituto Italiano di Tecnologia, Torino, Italy}

\journal{Journal Name}

\begin{document}

\begin{frontmatter}

\begin{abstract}
A fungal skin is a thin flexible sheet of a living homogeneous mycelium made by a filamentous fungus. The skin could be used in future living architectures of adaptive buildings and as a sensing living skin for soft self-growing/adaptive robots. In experimental laboratory studies we demonstrate that the fungal skin is capable for recognising mechanical and optical stimulation. The skin reacts differently to loading of a weight, removal of the weight, and switching illumination on and off. These are the first experimental evidences that fungal materials can be used not only as mechanical `skeletons' in architecture and robotics but also as intelligent skins capable for recognition of external stimuli and sensorial fusion.
\end{abstract}


\end{frontmatter}

\section{Introduction}

Flexible electronics, especially electronic skins~\cite{soni2020soft,ma2017self,zhao2017electronic} is amongst the most rapidly growing and promising fields of novel and emergent hardware. The electronic skins are made of flexible materials where electronics capable of tactile sensing~\cite{chou2015chameleon,yang2015tactile,wang2015recent,pu2017ultrastretchable} are embedded. The electronic skins are capable of low level perception~\cite{chortos2016pursuing,park2014stretchable} and could be developed as autonomous adaptive devices~\cite{nunez2019energy}. Typical designs of electronic skins include thin-film transistor and pressure sensors integrated in a plastic substrate~\cite{wang2013user}, micro‐patterned polydimethylsiloxane with carbon nanotube  ultra-thin films~\cite{wang2014silk,sekitani2012stretchable}, a large-area film synthesised by sulfurisation of a tungsten film~\cite{guo2017transparent}, multilayered graphene~\cite{qiao2018multilayer}, platinum ribbons~\cite{zhao2017electronic}, Polyethylene terephthalate (PET) based silver electrodes~\cite{zhao2015flexible}, digitally printed hybrid electrodes for electromyographic recording ~\cite{scalisi2015} or for piezoresistive pressure sensing ~\cite{chiolerio2014}, or channels filled with conductive polymer~\cite{chiolerio2020tactile}.

Whilst the existing designs and implementations are highly impactful, the prototypes of electronic skins lack a capacity to self-repair and grow. Such properties are useful, and could be necessary, when an electronic skin is used in e.g. unconventional living architecture~\cite{adamatzky2019fungal}, soft and self-growing robots~\cite{el2018development,sadeghi2017toward,rieffel2014growing,greer2019soft} and development of intelligent materials from fungi~\cite{Meyer2020eurofung, haneef2017advanced,Jones2020composites, Wosten2019MMs}. Based on our previous experience with designing tactile, colour sensors from slime mould \emph{Physarum polycephalum}~\cite{adamatzky2013towards, adamatzky2013slime, whiting2014towards} and our recent results on fungal electrical activity~\cite{adamatzky2018spiking,beasley2020capacitive,beasley2020mem} we decided to propose a thin layer of homogeneous mycelium of the trimitic polypore species \emph{Ganoderma resinaceum} as a live electronic skin and thus investigate its potential to sense and respond to tactile and optical stimuli. We call the fungal substrate, used in present paper, `fungal skin' due to its overall appearance and physical feeling. In fact, several species of fungi have been proposed as literal skin substitutes and tested in wound healing~\cite{Hamlyn1991,Hamlyn1994,su1997fungal,su1999development,xu2019wound,Narayanan2020scaffold}.

The paper is structured as follows. We introduce the protocol for growing the fungal skin and the methods of electrical activity recording in Sect.~\ref{methods}. Patterns of electrical activity of the fungal skin are analysed in Sect.~\ref{results}. Results are considered in a wider context and directions of future studies are outlined in Sect.~\ref{discussion}.

\section{Methods}
\label{methods}

 Potato dextrose agar (PDA), malt extract agar (MEA) and malt extract (ME) were purchased from Sigma-Aldrich (USA). The \emph{Ganoderma resinaceum} culture used in this experiment was obtained from a wild basidiocarp found at the shores of \emph{Lago di Varese}, Lombardy (Italy) in 2018 and maintained in alternate PDA and MEA slants at MOGU S.r.l. for the last 3 years at 4~ºC under the collection code 019-18. 

\begin{figure}[!tbp]
    \centering
        \subfigure[]{\includegraphics[width=0.4\textwidth]{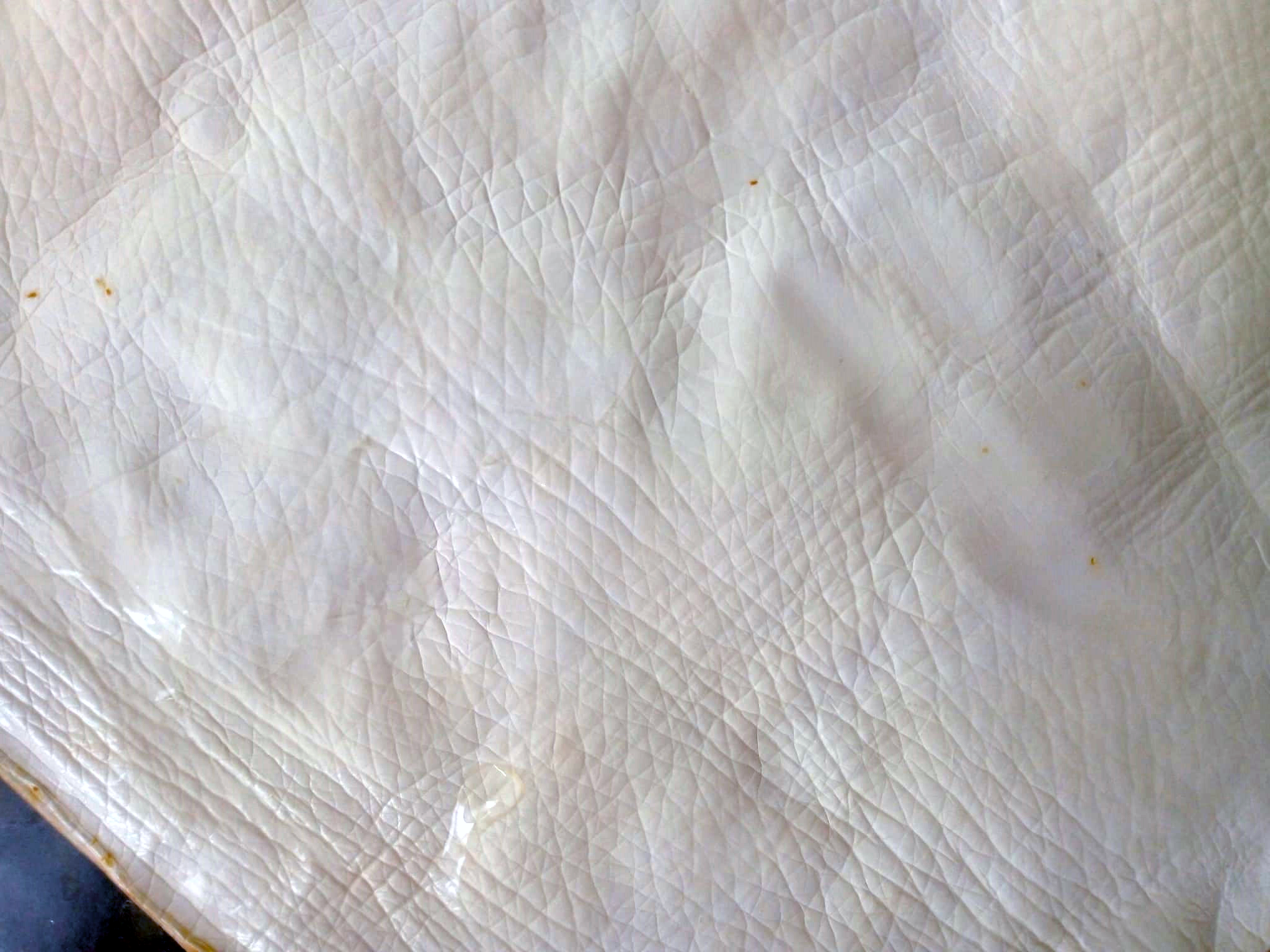}}
        \subfigure[]{\includegraphics[width=0.4\textwidth]{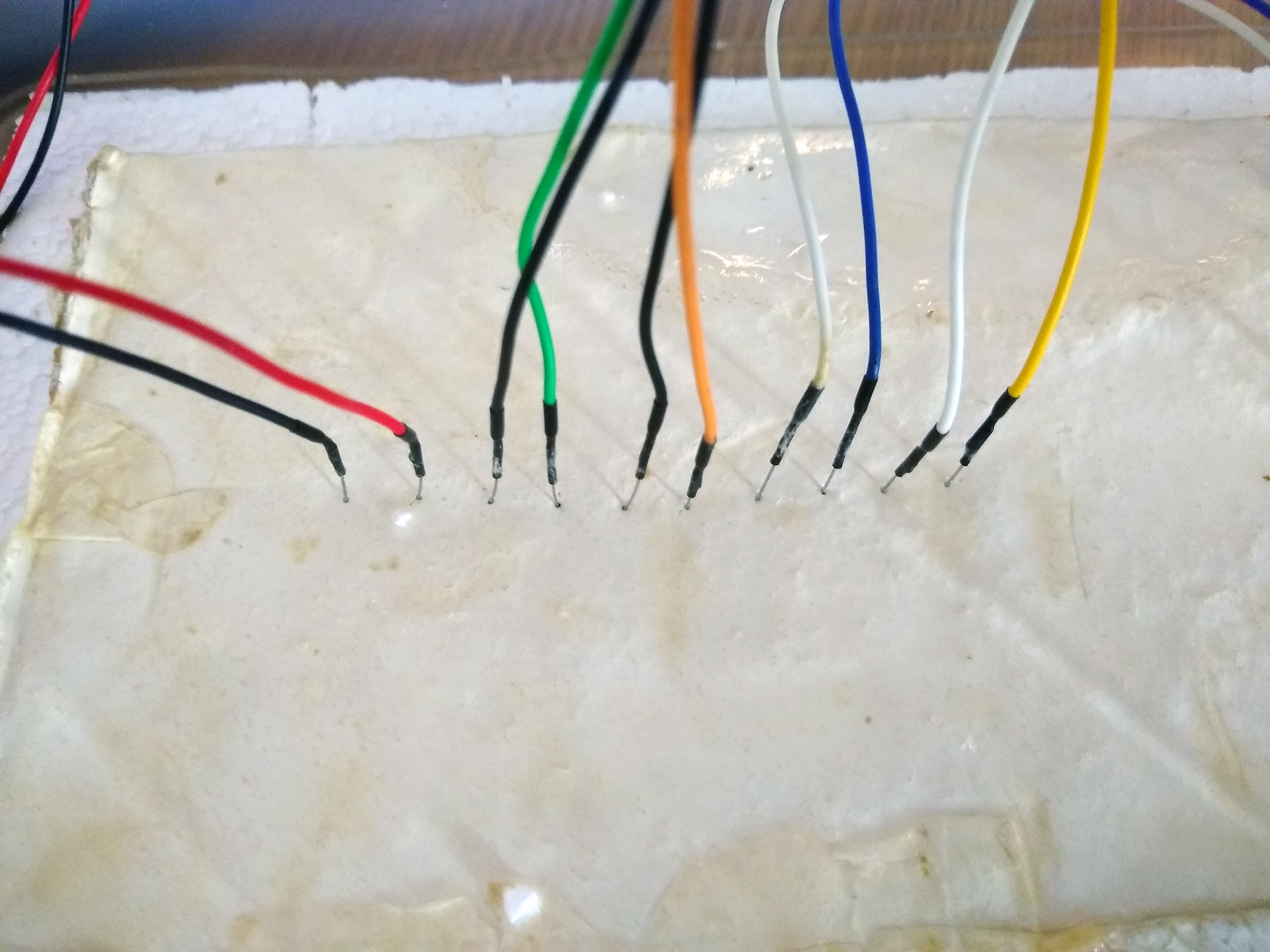}}
        \subfigure[]{\includegraphics[width=0.8\textwidth]{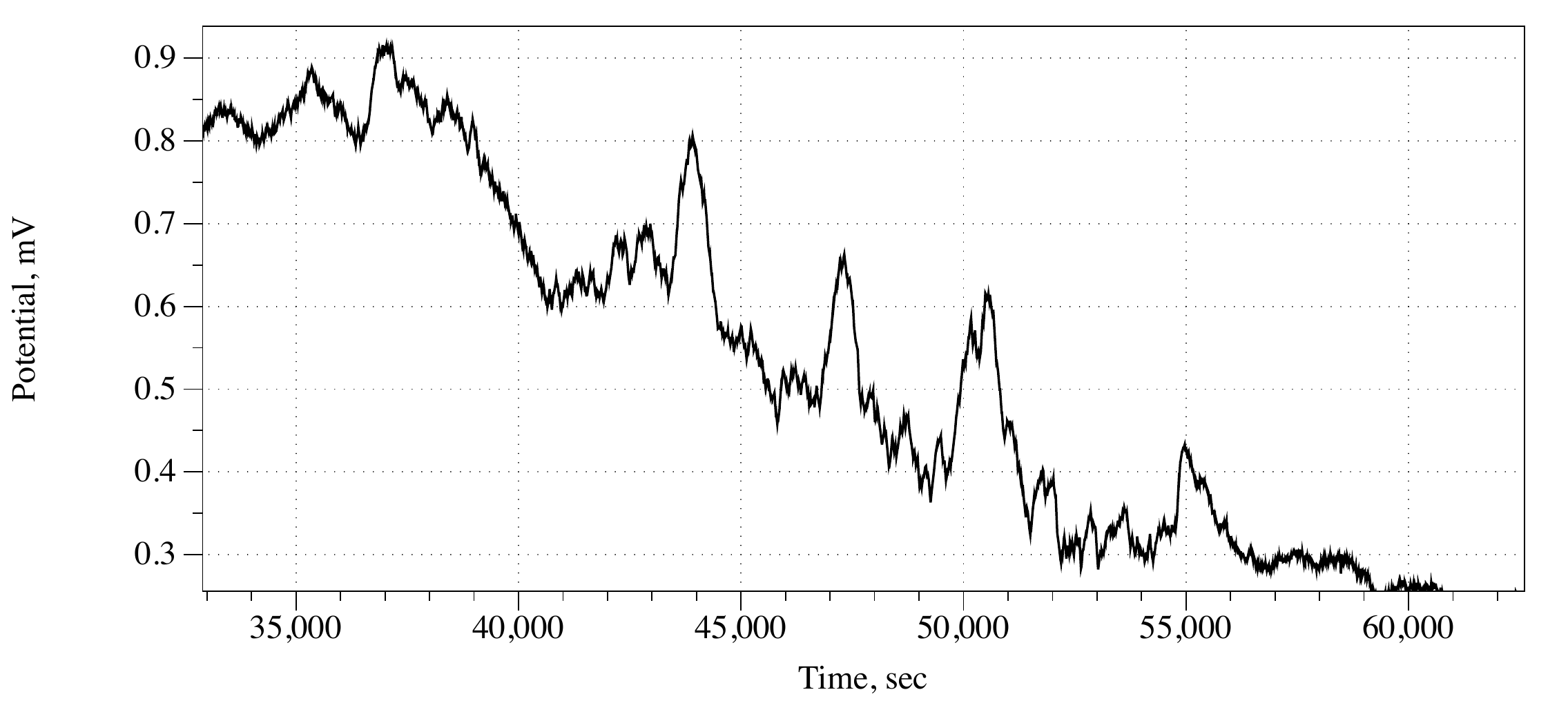}}
        \subfigure[]{\includegraphics[width=0.8\textwidth]{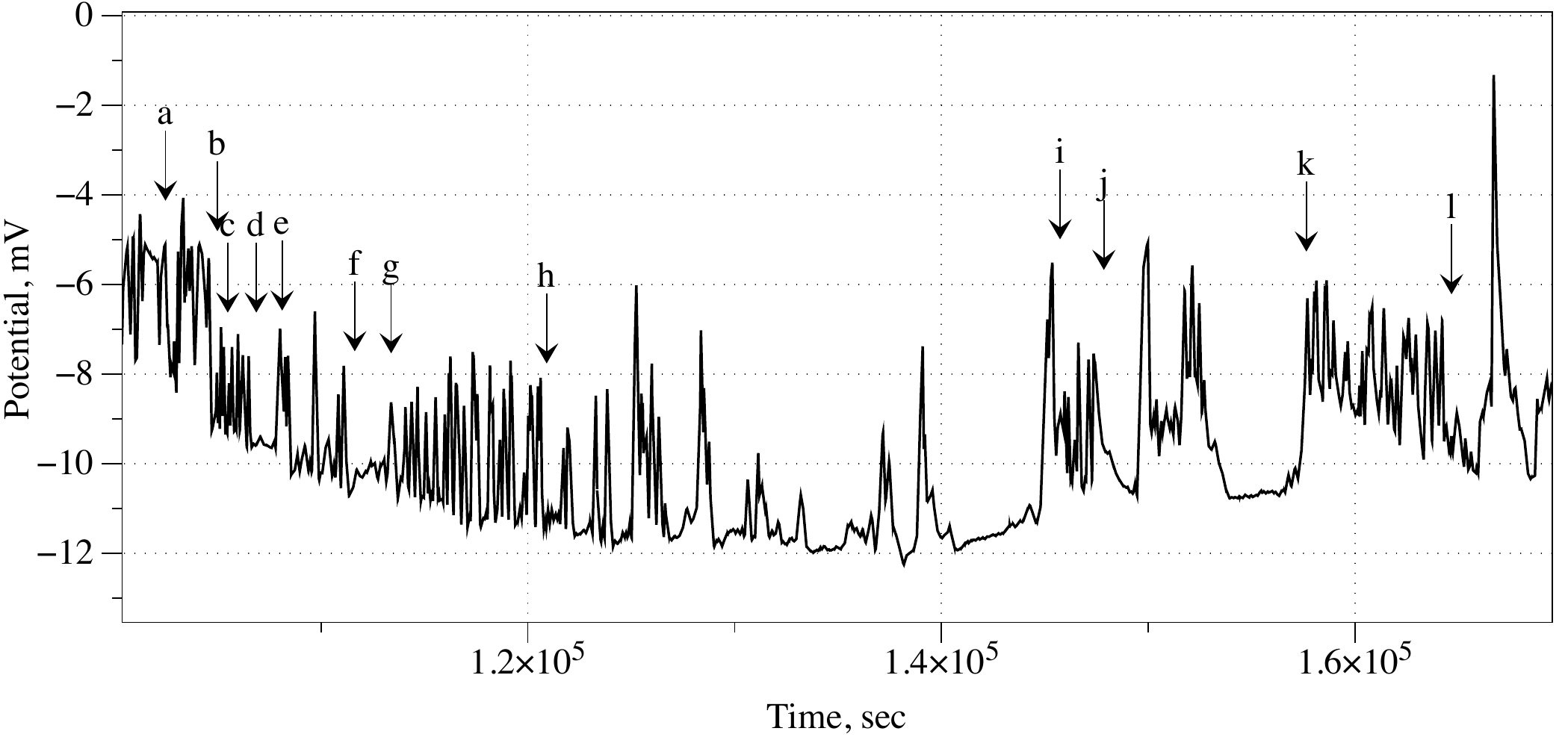}}
    \caption{Recording of electrical activity of fungal skin. 
    (a)~Close-up texture detail of a fungal skin.
    (b)~A photograph of electrodes inserted into the fungal skin.
  (c)~Train of three low-frequency spikes, average spike's width there is 1500~sec, a distance between spike peaks is 3000~sec and average amplitude is 0.2~mV.
    (d)~Example of several train of high-frequency spikes. Each train $T_{xy}=(A_{xy}, W_{xy}, P_{xy})$ is characterised by average amplitude of spikes $A_{xy}$~mV, width of spikes $W_{xy}$~sec and average distance between neighbouring spikes' peaks $P_{xy}$~sec: 
    $T_{ab}=(2.6, 245, 300)$,
    $T_{cd}=(1.7, 160, 220)$,
    $T_{ef}=(1.6, 340, 340)$,
    $T_{gh}=(2.5, 240, 350)$,
    $T_{ij}=(2.5, 220, 590)$,
    $T_{kl}=(2.6, 290, 440)$.  
    }
    \label{fig:electrodes}
\end{figure}

The fungal skin was prepared as follows. \emph{G. resinaceum} was grown on MEA plates and a healthy mycelium plug was inoculated into an Erlenmeyer flask containing 200~ml of 2\% ME broth (MEB) that was then incubated under continuous shaking at 200~rpm and 28~\textsuperscript{o}C for 5 days. Subsequently, this liquid culture was homogenised for 1 minute at max. speed in a sterile 1L Waring laboratory blender (USA) containing 400~mL of fresh MEB, the resulting 600~mL of living slurry were then poured into a 35 by 35~cm static fermentation tray. The slurry was let to incubate undisturbed for 15 days to allow the fungal hyphae to inter-mesh and form a floating mat or skin of fungal mycelium. Finally, a living fungal skin c. 1.5~mm thick was harvested (see texture of the skin in Fig.~\ref{fig:electrodes}a), washed in sterile demineralised water, cut to the size 23~cm by 11~cm and placed onto a polyurethane base to keep electrodes stable during the electrical characterisation steps (Fig.~\ref{fig:electrodes}b).

The electrical activity of the skin was measured as follows. We used iridium-coated stainless steel sub-dermal needle electrodes (Spes Medica S.r.l., Italy), with twisted cables. The pairs of electrode were inserted in fungal skin (Fig.~\ref{fig:electrodes}b). In each pair we recorded a difference in electrical potential between the electrodes. We used ADC-24 (Pico Technology, UK) high-resolution data logger with a 24-bit Analog to Digital converter, galvanic isolation and software-selectable sample rates. We recorded electrical activity with a frequency of one sample per second. We set the acquisition voltage range to 156~mV with an offset accuracy of 9~$\mu$V to maintain a gain error of 0.1\%. For mechanical stimulation with 30~g nylon cylinder, contact area with the fungal skin was c.~35~mm disc. For optical stimulation we used an aquarium light, array of LEDs, 36 white LEDs and 12 blue LEDs, 18~W, illumination on the fungal skin was 0.3~LUX.

\section{Results}
\label{results}

Endogenous electrical activity of the fungal material is polymorphic. Low  and high frequency oscillations patterns can emerge intermittently. A train of four spikes in Fig.~\ref{fig:electrodes}c is an example of low frequency oscillations. Trains of high-frequency spikes are exemplified in Fig.~\ref{fig:electrodes}d.

\begin{figure}[!tbp]
    \subfigure[]{
    \includegraphics[width=1\textwidth]{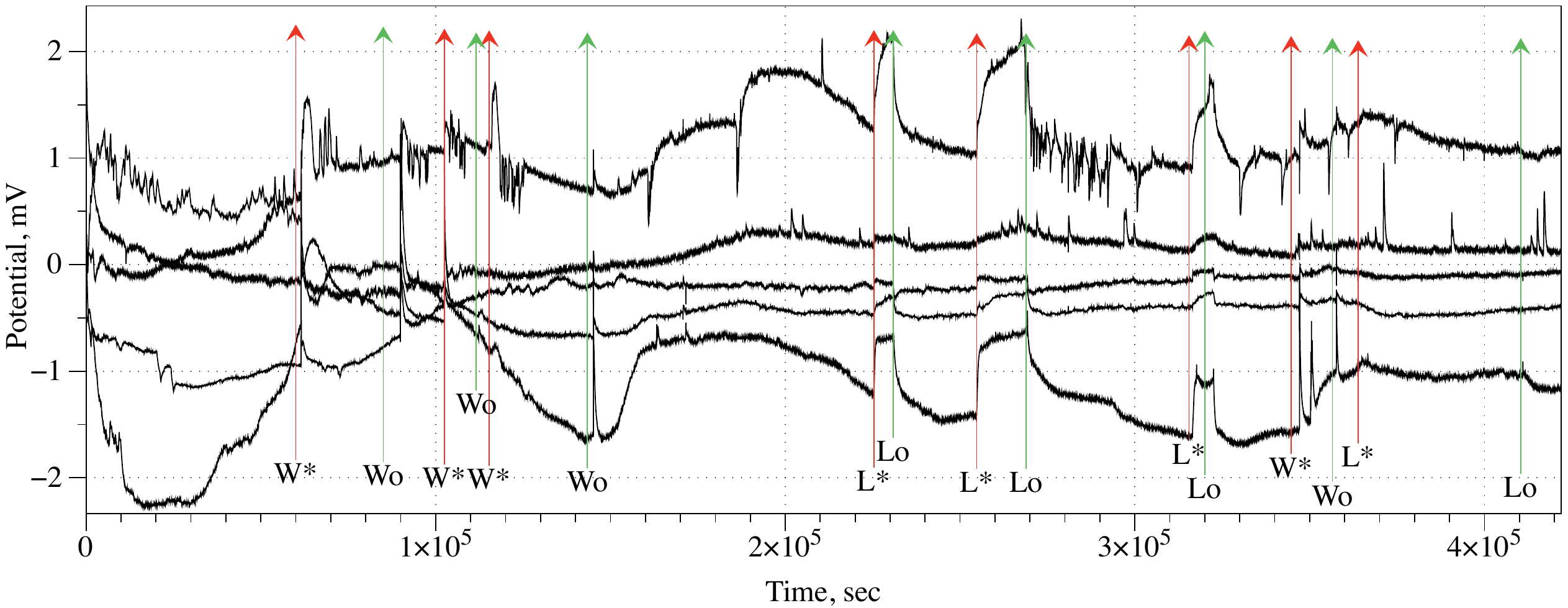}
    }
    \subfigure[]{
    \includegraphics[width=0.75\textwidth]{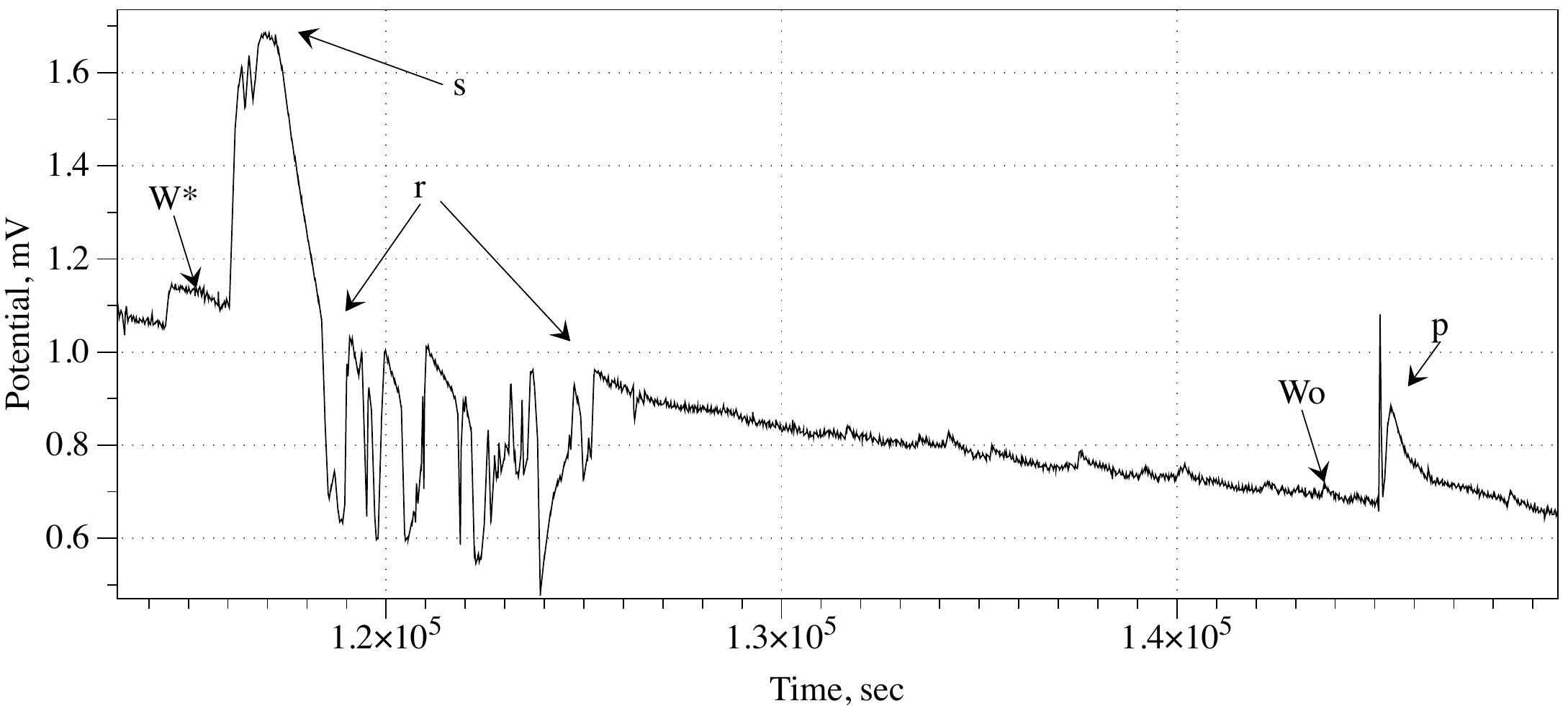}
    }
    \subfigure[]{
    \includegraphics[width=0.48\textwidth]{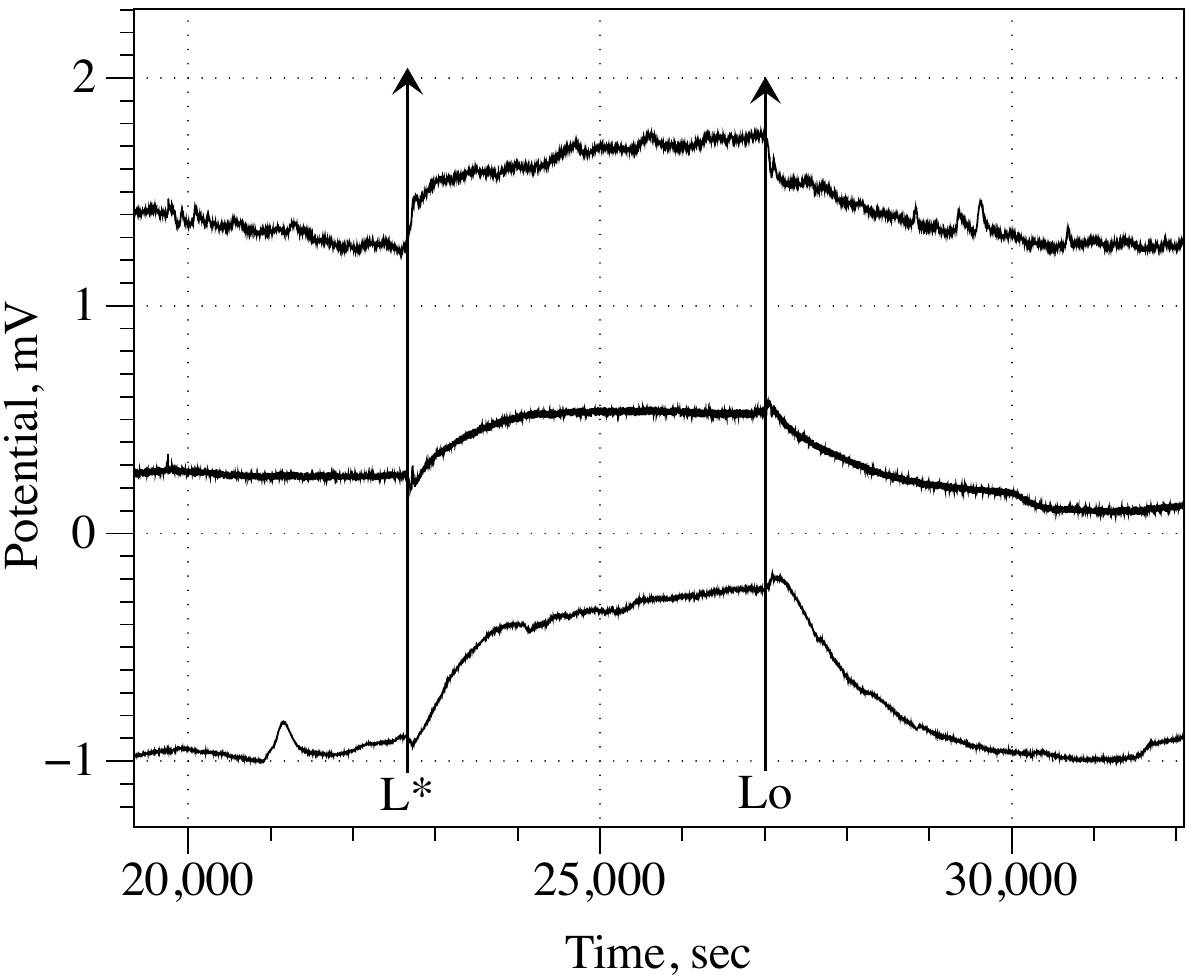}
    }
    \subfigure[]{
    \includegraphics[width=0.48\textwidth]{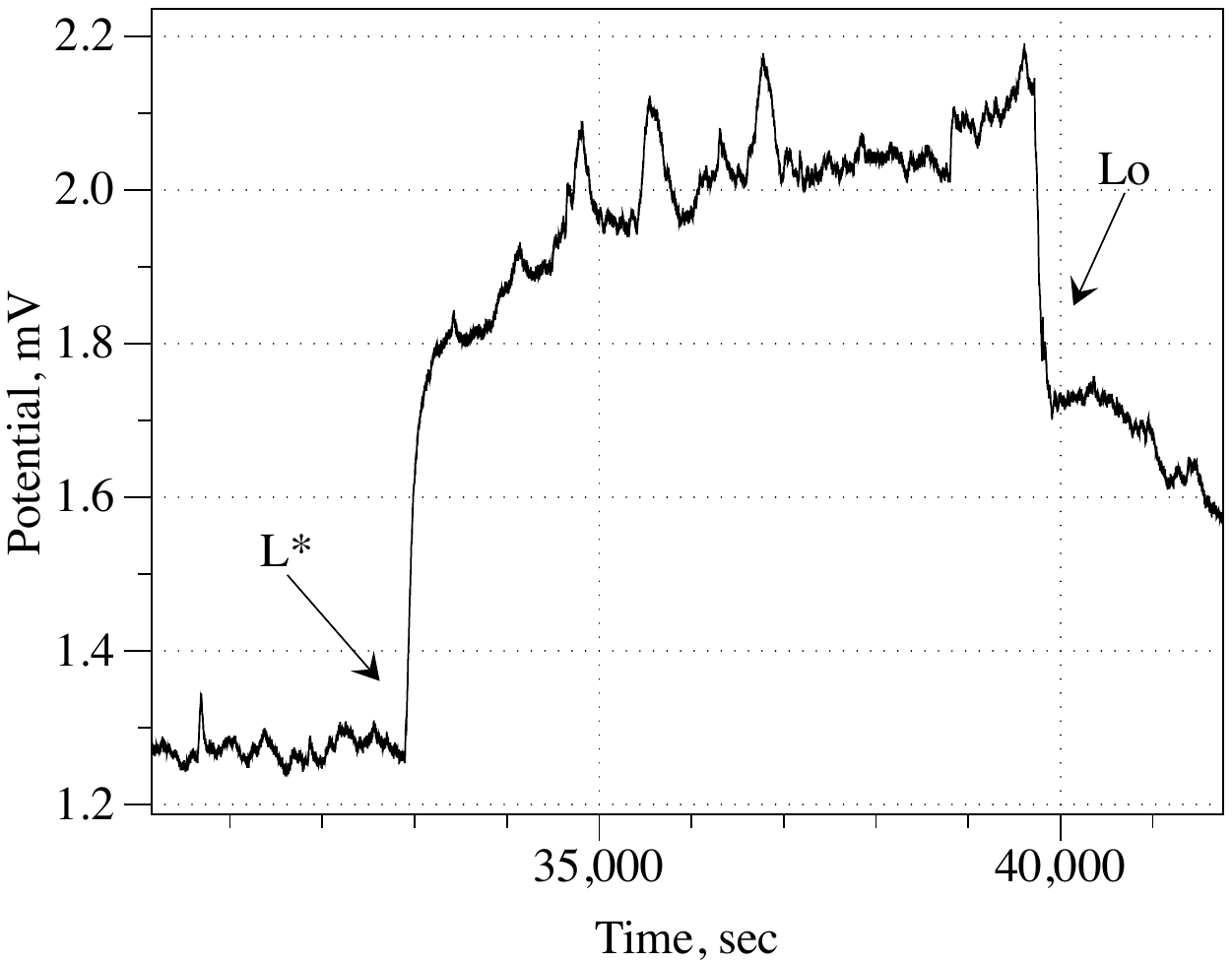}
    }
    \caption{Fungal skin response to mechanical and optical stimulation.
    (a)~Exemplar recording of fungal skin electrical activity under tactile and optical stimulation. Moments of applying and removing a weight are shown as `W*' and `Wo' and switching light ON and OFF as `L*' and `Lo'.
    (b)~Exemplar response to mechanical stimulation. Moments of applying and removing a weight are shown as `W*' and `Wo'. High-amplitude response is labelled `s'. This response is following by a train of spikes `r'. A response to the removal of the weight is labelled `p'.
    (c)~Exemplar response of fungal skin to illumination, recorded on three pairs of differential electrodes. `L*' indicates illumination is applied, `Lo' illumination is switched off. 
    (d)~A train of spikes on the raised potential as a response to illumination.
    }
    \label{fig:examplarresponses}
\end{figure}

Electrical responses to tactile loading and illumination are distinctive. An example of several rounds of stimulation is shown in Fig.~\ref{fig:examplarresponses}a. 
The fungal skin responds to loading of a weight with a high-amplitude wide spike of electrical potential sometimes followed by a train of high-frequency spikes. The skin also responds to removal of the weight by a high-amplitude spike of electrical potential.

An exemplar response to loading and removal of weight is shown in Fig.~\ref{fig:examplarresponses}b. The parameters of the fungal skin responses to the weight being placed on the skin are the following. An average delay of the response (the time from weight application to a peak of the high-amplitude spike) is 911.4~sec ($\sigma$=1280.1, minimum 25~sec and maximum 3200~sec). An average amplitude of the response spike (marked `s' in the example Fig.~\ref{fig:examplarresponses}b) is 0.4~mV ($\sigma$=0.2, minimum 0.1~mV and maximum 0.8~mV). An average width of the response spike is 1261.8~sec ($\sigma$=1420.3, minimum 199~sec and maximum 4080~sec), meaning that the average energy consumed per current unit, associated to the response, is approximately 0.5 J/A. A train of spikes (marked `r' in the example Fig.~\ref{fig:examplarresponses}b), if any, following the response spike usually has 4 or 5 spikes. The fungal skin responds to removal of the weight (the response is marked `p' in the example Fig.~\ref{fig:examplarresponses}b) with a spike which average amplitude is 0.4~mV ($\sigma$=0.2, minimum 0.2~mV and maximum 0.85~mV). Amplitudes are less indicative than frequencies because an amplitude depends on the position of electrodes with regards to propagating wave of excitation. An average width of the spike is 774~sec ($\sigma$=733.1, minimum 100~sec and maximum 2000~sec. A response of the fungal skin to removal of the weight was not observed in c. 20\% of differential electrode pairs. The average response time is 385.5~sec ($\sigma$=693.3~sec, minimum 77~sec and maximum 1800~sec).

Fungal skin's response to illumination is manifested in the raising of the baseline potential, as illustrated in the exemplar recordings in Fig.~\ref{fig:examplarresponses}c. 
In contrast to mechanical stimulation response the response-to-illumination spike does not subside but the electrical potential stays raised till illumination is switched off. An average amplitude of the response is 0.61~mV ($\sigma$=0.27, minimum 0.2~mV and maximum 1~mV). The raise of the potential starts immediately after the illumination is switched on. The potential reaches its maximum and goes onto plateau in 2960~sec in average ($\sigma$=2201, minimum 879~sec and maximum 9530~sec). Typically, we did not observe any spike trains after the illumination switched off however in a couple of trials we witnessed spike trains on top of the raised potential, as shown in Fig.~\ref{fig:examplarresponses}d.

\section{Discussion}
\label{discussion}

\begin{figure}[!tbp]
    \subfigure[]{
    \includegraphics[width=0.49\textwidth]{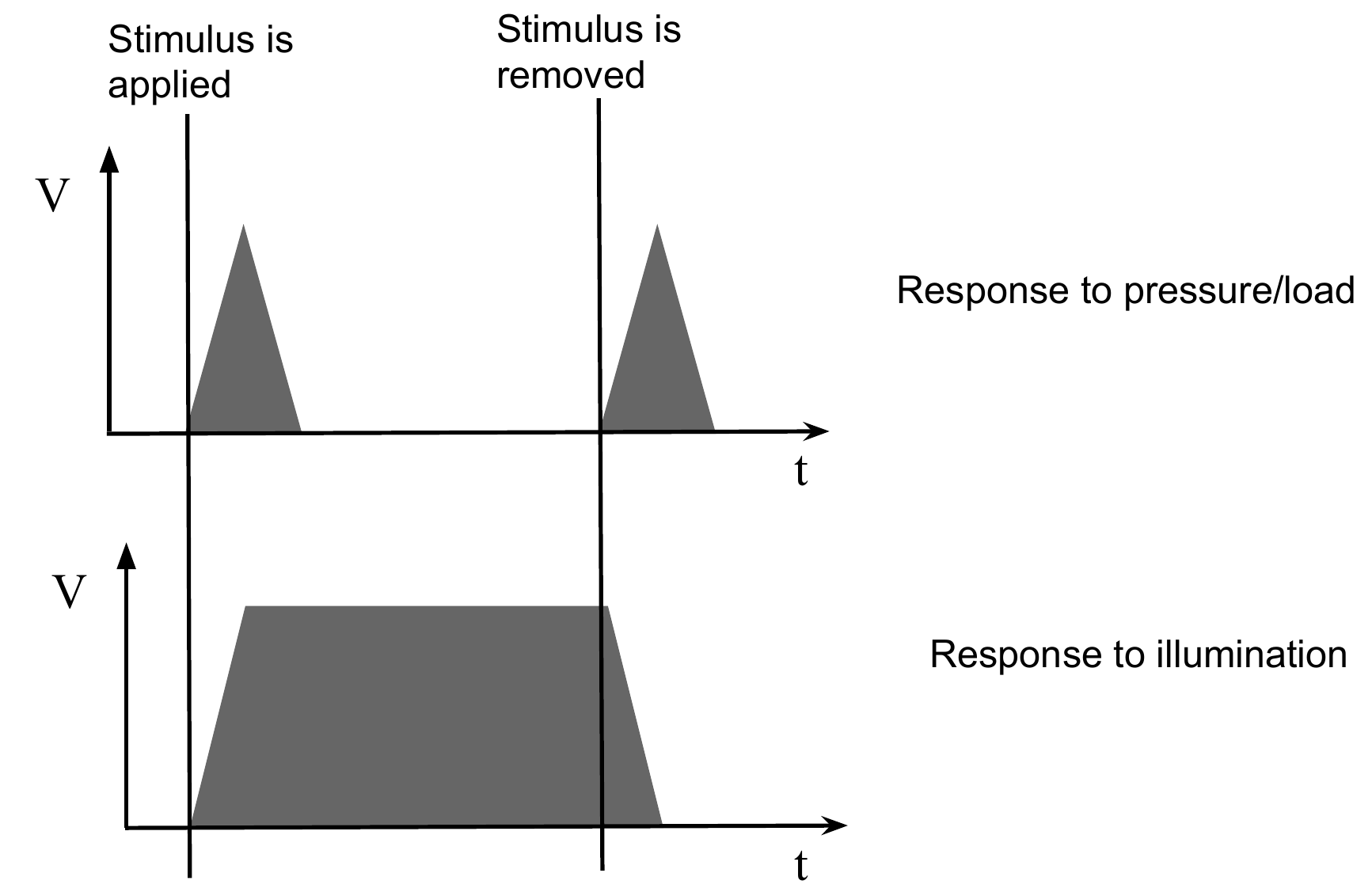}
    }
    \subfigure[]{
    \includegraphics[width=0.49\textwidth]{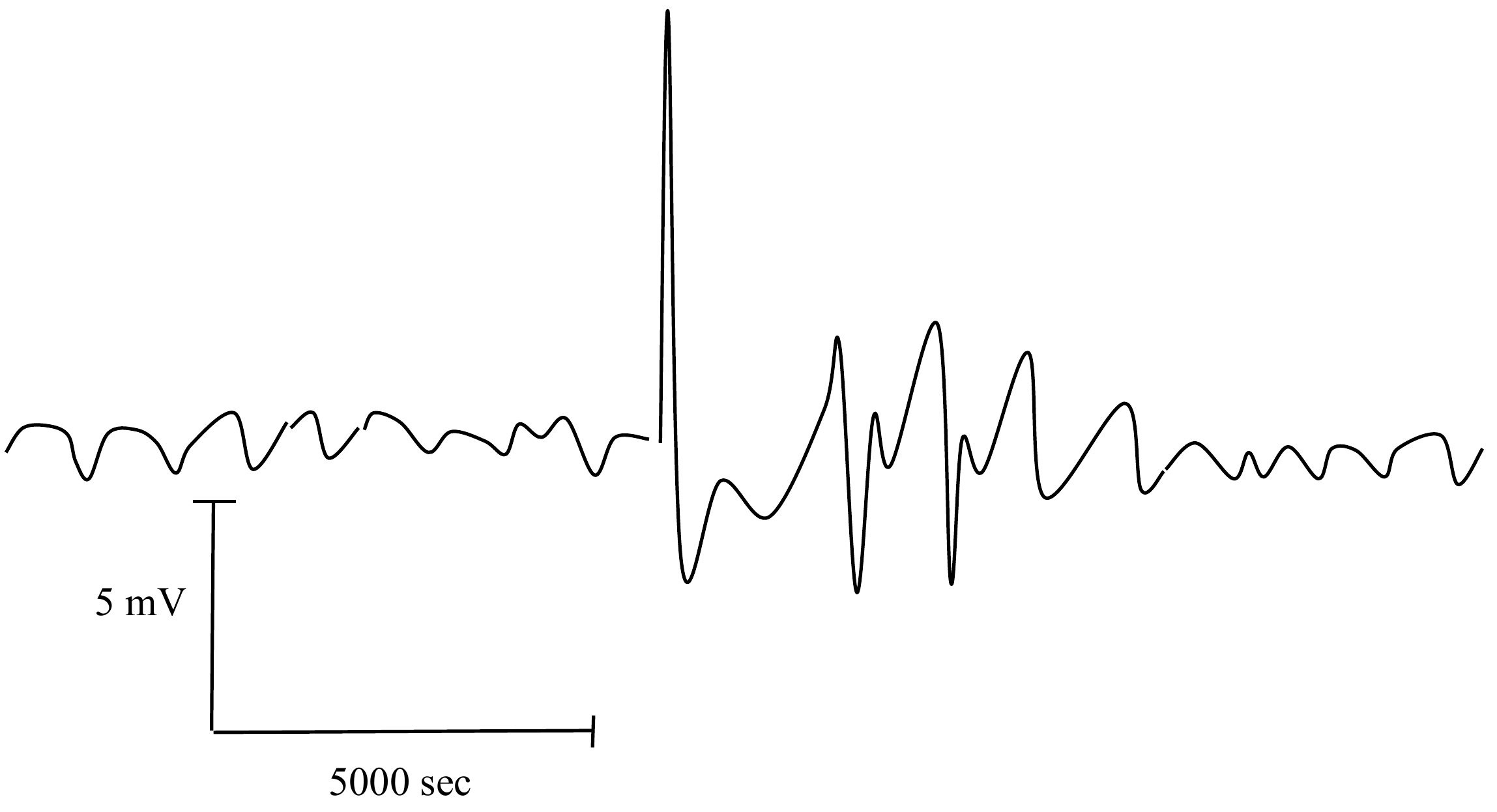}
    }
    \caption{
    (a)~A scheme of the fungal skin responses to mechanical load and optical stimulations.
    (b)~Slime mould \emph{P. polycephalum} response to application of 0.01~g glass capillary tube. Redrawn from \cite{adamatzky2013slime}.}
    \label{fig:discussion}
\end{figure}

We demonstrated that a thin sheet of homogeneous living mycelium of \emph{Ganoderma resinaceum}, which we named `fungal skin', shows pronounced electrical responses to mechanical and optical stimulation. Can we differentiate between the fungal skin's response to mechanical and optical stimulation? Definitely, see Fig.~\ref{fig:discussion}a. The fungal skin responds to mechanical stimulation with a 15~min spike of electrical potential, which diminishes even if the applied pressure on the skin remains. The skin responds to optical stimulation by raising its electrical potential and keeping it raised till the light is switched off.

Can we differentiate the responses to loading and removal of the weight? Yes. Whilst amplitudes of `loading' and `removal' spikes are the same (0.4~mV in average) the fungal skin average reaction time to removal of the weight is 2.4 times shorter than the reaction to loading of the weight (385~sec versus 911~sec). Also `loading' spikes are 1.6 times wider than `removal' spikes (1261~sec versus 774~sec).

Fungal skin response to weight application is, in some cases, esp. Fig.~\ref{fig:examplarresponses}y, similar to response of slime mould to application of the light weight~\cite{adamatzky2013slime}. The following events are observed (Fig.~\ref{fig:discussion})b: oscillatory activity before stimulation, immediate response to stimulation, prolonged response to stimulation as a train of high-amplitude spikes, return to normal oscillatory activity. This might indicate some universal principles of sensing and information processing in fungi and slime moulds.

The sensing fungal skin proposed has a range of advantages comparing to other living sensing materials, e.g. slime mould sensors~\cite{adamatzky2013towards,adamatzky2013slime,whiting2014towards} 
electronic sensors with living cell components~\cite{kovacs2003electronic},
 chemical sensors using living taste, olfactory, and neural cells and tissues~\cite{wu2015bioanalytical} and tactile sensor from living cell culture~\cite{minzan2013toward}. The advantages are low production costs, simple maintenance, durability. The last but not least advantage is scalability: a fungal skin patch can be few microns and it can be grown to several metres in size.

In future studies we will aim to answer the following questions. Would it be possible to infer a weight of the load applied to the fungal skin from patterns of its electrical activity? Would the fungal skin indicate directionality of the load movement by its spiking activity? Would it be possible to locate the position of the weight within the fungal network? Would it be possible to map a spectrum of the light applied to the skin onto patterns of the skin's electrical activity?

\section*{Acknowledgement}

This project has received funding from the European Union's Horizon 2020 research and innovation programme FET OPEN ``Challenging current thinking'' under grant agreement No 858132.


\end{document}